\documentclass[11pt]{article}

\usepackage[english,francais]{babel}

\usepackage[latin1]{inputenc}
\usepackage{epsfig,graphicx}
\usepackage{latexsym}

\newfont{\ensmathquatorze}{msbm10 scaled 1400}
\newfont{\ensmathonze}{msbm10 scaled 1100}
\newfont{\ensmathdix}{msbm10}
\newfont{\ensmathneuf}{msbm10 scaled 833}
\newfont{\ensmathhuit}{msbm10 scaled 694}
\newfam\ensmathfam                        
\textfont\ensmathfam=\ensmathonze        
\scriptfont\ensmathfam=\ensmathdix       
\scriptscriptfont\ensmathfam=\ensmathhuit

\def\be{\begin{equation}}
\def\ee{\end{equation}}
\def\bea{\begin{eqnarray}}
\def\eea{\end{eqnarray}}

\def\d{{\rm d}}

\begin{document}

\selectlanguage{english}

\title{  Brownian Motion in wedges, last passage time and the
second arc-sine law  }

\author{ Alain Comtet$^{1,2}$ and Jean Desbois $^1$}

\maketitle	

{\small
\noindent
$^1$ Laboratoire de Physique Th\'eorique et Mod\`eles Statistiques.
Universit\'e Paris-Sud, B\^at. 100, F-91405 Orsay Cedex, France.

}
{\small
\noindent
$^2$ Institut Henri Poincar\'e, 11 rue Pierre et Marie Curie, 75005, Paris
}

\vskip1.5cm

\begin{abstract}
 We consider a planar Brownian motion starting from $O$ at time 
 $t=0$ and stopped
 at $t=1$ and a set $F= \{ OI_i \; ; \ i=1,2,\ldots ,n\}$ of $n$ semi-infinite
 straight lines emanating from $O$. Denoting by $g$ the last time when
 $F$ is reached by the Brownian motion, we compute the probability
 law of $g$. In particular, we show that, for a symmetric $F$ and
 even $n$ values, this law can be expressed as a sum of \ $\arcsin $ \
 or \  $(\arcsin )^2 $ \ functions. The 
 original result of Levy  is recovered as the
 special case $n=2$. A relation with the problem of reaction-diffusion of a
 set of three particles  in one dimension is discussed.
\end{abstract}

\vskip1.5cm

The first arc-sine law gives the distribution of the number of positive
partial sums in a sequence of independent and identically distributed
random variables. It was first discovered by P. Levy in his study of the
linear Brownian motion and then discussed a lot for its relevance to the
coin-tossing game \cite{feller}. The second arc-sine law, also discovered by
P. Levy \cite{levy},
provides an information on the last passage time which can be stated as
follows. Consider a linear Brownian motion $B(\tau)$ starting at $0$ at
time  $\tau=0$ and stopped at time $t$ and let $g$ be the last time when 
$0$ is visited.
The random variable
\be\label{a1}
g= \sup \{\ {\tau <t,B(\tau)=0}\}
\ee
satisfies

\bea
P(g<u)     &=& \frac{2}{\pi } \arcsin \sqrt{ \frac{u}{t}}        
 \label{asl}  \\
 \mbox{with the density } \quad  {\cal P}(u) &=& \frac{1}{\pi }
  \frac{1}{ \sqrt{u(t-u)}   } \label{asl1} 
\eea 
 
\vskip.5cm
\noindent
Over the years, this result has been extended in several different directions 
(see for instance, \cite{yor,Y2}) and is still a subject of active research in
probability \cite{Y1}.
 Generalizations of the first arc-sine law have also
 been considered  in different  contexts
 (one-dimensional diffusion in a random medium
 \cite{MC},  Brownian motion  on graphs  \cite{JD} and, also,
  in two dimensions   \cite{bingham}).

\vskip.5cm


\noindent
The purpose of this Letter is to present a two dimensional generalization of 
the law
(\ref{asl}). As a  by-product of this result we also derive an explicit
 expression of the
 first passage time distribution which is  relevant for a problem of
 reaction-diffusion involving three identical particles. 
\noindent

Exit problems for Brownian motion have a rich history and several
applications in physics (see for instance \cite{RE}). They are in particular
 related with problems of
capture of independent Brownian particles diffusing on the line. This
connection was first anticipated by Arratia \cite {A} and then
discussed  in the 
mathematics \cite {NOC,B} and physics litterature \cite {RE,F,BA,M}
 mainly in the context of reaction-diffusion
 models. In the case of three particles, the process
 $(x_1(t)-x_2(t)$, 
$x_2(t)-x_3(t))$ defines a certain diffusion in a quadrant of $R^2$.
By a suitable transformation, this process can be mapped on a diffusion
inside a wedge whose angle depends on the diffusion  constants. Using this
correspondance, it has been shown that the first passage time through the
wedge gives the survival probability; a quantity which decays with a power
law which only depends on the angle of the wedge \cite {BA,DB}.
 At the end of this work, we exploit this correspondance to compute exactly
  (and not only asymptotically)  
the first collision time distribution for a three particle problem. Our
approach is based on an identity relating the first passage and last passage
distribution which has an interesting probabilistic interpretation\cite {Y3}.

\vskip.5cm
 To begin with, let us start by considering,
 as in  Figure $1$, a wedge of apex $O$ and angle $\phi$
   with a boundary $F= \{ OI_i \; ; \ i=1,2 \} $ and a two dimensional
   Brownian motion
   $\overrightarrow{r}(t) $
  starting from $O$ at $t=0$ and stopped at $t$ somewhere in the plane.
 We denote by $g$ the last time when $F$ is visited and compute
  the probability $P(g<u)$. Due to the scaling property of the
   Brownian motion, this distribution is a function of the reduced variable
    $u/t$. In the following, we will for simplicity set $t=1$.

\begin{figure}[!h]
\begin{center}
\includegraphics[scale=.4,angle=0]{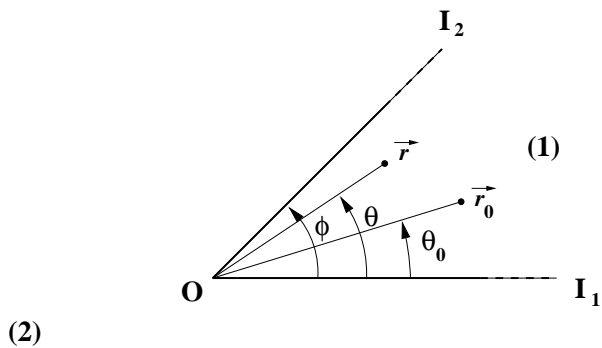}
\caption{  The frontier $F= \{ OI_1, OI_2\}$ of the  wedge divides the plane
into 2 regions, {\bf (1)} and {\bf (2)}.  }
\end{center}
\end{figure}

\vskip.5cm
\noindent
Suppose that the particle reaches some point $\overrightarrow{r_0}$ at time
 $t=u$ (see Figure 1).
  Clearly, if  $\overrightarrow{r_0}$ belongs to region {\bf (1)}
   (resp.{\bf (2)}), the particle must stay in {\bf (1)}
   (resp.{\bf (2)}) between  $t=u$ and  $t=1$ in order to satisfy the
   condition $g<u$. We can therefore write
  
\be\label{n1}  
 P(g<u) = P^{(1)}(u) +  P^{(2)}(u)
\ee

\vskip.3cm
\noindent

Expressing the fact that the propagation is free between  $t=0$ and  $t=u$
and that the particle has not hit the boundary between $t=u$ and $t=1$, we get
 
\be\label{n2}  
 P^{(i)}(u)=\int_{(i)} \d ^2 \overrightarrow{r_0}
\int_{(i)} \d ^2 \overrightarrow{r} \; \frac{1}{2\pi u} 
 \;  e^{- \frac{r_0^2}{2u}} \; G^{(i)}( \overrightarrow{r},1;
 \overrightarrow{r_0} ,u) \qquad ; \quad i=1,2
\ee
 
\vskip.2cm
\noindent
The propagator  $G^{(1)}$ satisfying the diffusion equation 
 with Dirichlet boundary conditions on $F$ is given by 
  
\be\label{n3}   
G^{(1)} = \frac{2}{\phi \; (1-u)} \; \sum_{m=1}^{\infty }
 \sin \frac{m\theta\pi }{\phi } \; \sin \frac{m\theta_0\pi }{\phi }
 \; e^{-\frac{r^2+r_0^2}{2(1-u)}} \; I_{\frac{m\pi }{\phi }} \left(
  \frac{rr_0}{1-u} \right)
\ee
where $I_{\nu }$  is a modified Bessel function and the notations are
defined on Figure 1.
 
\vskip.8cm
\noindent
Performing the spatial integrations in (\ref{n2}), we get \cite{GRAD}

\be\label{n4}  
P^{(1)}(u) = \frac{1}{\pi \; \phi} \; \sum_{p,k=0}^{\infty }
 u^{p\frac{\pi}{\phi} + k + \frac{\pi}{2 \phi} }
 \; \frac{ \left[ \Gamma (p\frac{\pi}{\phi} + k + \frac{\pi}{2 \phi} )
  \right]^2 }
 {\Gamma (\frac{2 p \pi}{\phi} + \frac{\pi}{ \phi} +k+1 )   }
  \; \frac{1}{k!}
\ee

$P^{(2)} $ is obtained by the change $\phi \to 2\pi - \phi$ in
 (\ref{n4}). Therefore, for arbitrary values of  $\phi $ 
  the law $P(g<u) $ is written in terms of a double series.

\vskip.3cm

\noindent
As a check, let us first consider the special case $\phi= \pi$. We may
write

\bea
P(g<u) = 2 P^{(1)}(u) &=& \frac{2}{\pi^2} \; u^{1/2} \sum_{p,k=0}^{\infty }
 \frac{u^{p+k}}{k!} \frac{ \left[ \Gamma (p+k+1/2)  \right]^2   }
 { (2p+k+1)!}   \label{n5}  \\ 
     &=&  \frac{2}{\pi } \arcsin \sqrt{u}            \label{n6} 
\eea

\noindent
The fact that one recovers Levy's second arc-sine law is not surprising since, when
  $\phi= \pi$, $F$ divides the plane into two half-planes. Therefore the
  component
   of the Brownian motion parallel to $F$ factorises and plays no role: we are thus left
   with a one dimensional  problem.
 
\vskip.5cm
\noindent
Coming back to  general  values of $\phi $, we can  derive the behavior of
 the probability density ${\cal P}$ $ ( \equiv \frac{\d P(g<u)}{\d u} ) $
 when $u\to 0$ and  $u\to 1$. By using (\ref{n1}) and (\ref{n4}), one gets
 a power-law behavior when $u\to 0^+$
  
\bea
 {\cal P}(u) &\sim& \frac{1}{\pi } \; u^{-1/2}
 \qquad \mbox{for } \  \phi = \pi 
 \label{n7}    \\ 
 {\cal P}(u) &\sim&    C(\mu ) \; u^{ \frac{\mu }{2} -1}              
 \qquad \mbox{for } \  \phi \ne \pi 
 \label{n8}    \\ 
 \mbox{with } \quad  C(\mu ) 
         &=& \frac{\mu^2}{2\pi^2} \; 
  \frac{ \left[ \Gamma(\frac{\mu }{2})    \right]^2  }{\Gamma (\mu +1)} 
  \label{n80}    \\ 
\mbox{and } \quad   \mu      &=&  \frac{\pi }{ 2\pi - \phi} \qquad \mbox{when}
 \quad 0< \phi < \pi 
\label{n9}    \\ 
  \mu    &=&    \frac{\pi }{  \phi} \qquad \mbox{when}
 \quad    \pi < \phi < 2 \pi 
  \label{n10}    
\eea

\vskip.5cm
\noindent
Now, for the limit $u \to 1^- $, using asymptotic expansions for  $\Gamma $ 
 functions and also an equivalence between series and integrals, we get
 
\be\label{n11}  
{\cal P}(u) \sim  \frac{1 }{  \pi} \; \frac{1 }{ \sqrt{1-u}} 
\ee

\vskip.2cm
\noindent
i.e. the same behavior as for (\ref{asl1}). The expression (\ref{n11})
 doesn't depend on $\phi$ and we have already seen that $\phi =\pi$ gives
  the Levy's law. Remark that $u \to 1^- $ corresponds to Brownian curves
   that stop close to $F$.
    Therefore, between $t=u$ and  $t=1$, the Brownian particle
 only ``sees'' an infinitesimal part of $F$, i.e. a straight line as for
  $\phi =\pi$. This is, in our opinion, why the result (\ref{n11})
doesn't depend on $\phi$. Actually, it only depends on the fact that the 
plane is divided by $F$ into two regions. We will come back to this point 
 latter on.
 
\vskip.5cm
\noindent

\begin{figure}[!h]
\begin{center}
\includegraphics[scale=.4,angle=0]{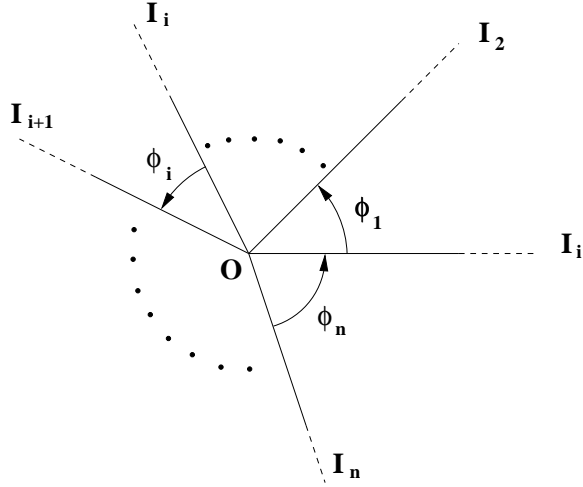}
\caption{ $F$ consists in $n$ semi-infinite straight lines starting from $O$. }
\end{center}
\end{figure}

\vskip.5cm
\noindent 
To go further, let us remark that for    $F= \{ OI_i \; ; \  i=1,2,\ldots ,n\}$
 as in Figure 2, (\ref{n1}) becomes simply:
 
\be\label{r1} 
 P(g<u) = \sum_{i=1}^n  \; P^{(i)}(u)
\ee 

\vskip.3cm
\noindent
(Replace $\phi $ by $\phi_i $ in (\ref{n4}) in order to get $P^{(i)}$).
 
\vskip.7cm
\noindent
Let us now specialize to the situation when $F$ is symmetric and $n$ is even
 ($n \equiv 2l$). In that case, $F$ consists in $l$ infinite straight lines
  crossing at point $O$ and dividing the plane into $2l$ equal angular
  sectors, each one of angle $\phi = \pi /l$.
 
\begin{figure}[!h]
\begin{center}
\includegraphics[scale=.4,angle=0]{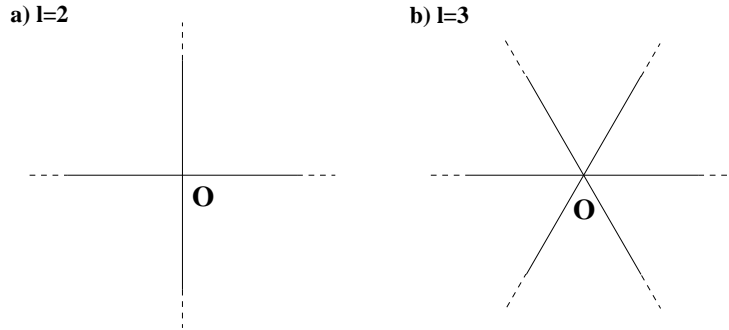}
\caption{ $F$ is symmetric.  The analytic form of $P(g<u)$
 will depend on the parity of $l$. Thus, it will be different for the cases a)
 and b). For further explanations, see text.}
\end{center}
\end{figure}

\vskip.3cm
\noindent
 Equation (\ref{r1}) writes

\be\label{r2}
P(g<u) \equiv P_l(u) =
 \frac{2}{\pi^2} \; l^2 \; u^{l/2} \sum_{p=0}^{\infty } u^{lp}
 \sum_{k=0}^{\infty }
 \frac{ \left[ \Gamma (lp+k+l/2)  \right]^2   }
 { (2lp+l+k)!}   
\frac{u^{k}}{k!} 
\ee

\vskip.6cm
\noindent
$l$ being an integer, we can sum the series and, finally,  get

\bea
 P_l (u) &=& \frac{2l }{\pi } \left(
  \sum_{k=0}^{l-1} \arcsin \left( \sqrt{u} \cos \frac{2\pi k}{l} \right) 
  \right)  \qquad , \quad  l \mbox{  odd} \label{r3}    \\
 P_l (u) &=& \frac{2l }{\pi ^2} \left(
  \sum_{k=0}^{l-1} \; (-1)^k  \;
 \left( \arcsin \left( \sqrt{u} \cos \frac{\pi k}{l} \right) \right)^2 
  \right)  \qquad , \quad  l \mbox{  even} \label{r4}  
\eea
which is the central result of this paper.

\vskip.3cm
\noindent
We remark that the correct small $u$ behavior for $P_l(u)$ follows from the two
identities 
\bea
    \sum_{k=0}^{l-1} \left( \cos \frac{2\pi k}{l} \right)^m &=& 0
  \qquad , \quad  l \mbox{ odd } \quad , \quad m=1,3,\ldots , l-2  \\
    \sum_{k=0}^{l-1} (-1)^k \; \left( \cos \frac{\pi k}{l} \right)^m &=& 0
  \qquad , \quad  l \mbox{  even } \quad , \quad m=0,2,4,\ldots , l-2
\eea

\vskip.6cm
\noindent
In particular

\bea
      P_1 (u)  &=&   \frac{2}{\pi } \arcsin \sqrt{u}           \label{r5}   \\
      P_2 (u)&=&    \frac{4}{\pi ^2 } \left( \arcsin \sqrt{u} \right)^2 = 
      P_1^2    \label{r6}   \\
      P_3 (u)&=&   \frac{6}{\pi } \left( 
      \arcsin \sqrt{u} \;   - \; 2 \; 
      \arcsin \frac{\sqrt{u}}{2}   \right)     \label{r7}   \\
      P_4 (u)&=&    \frac{8}{\pi ^2 }\left(
      \left( \arcsin \sqrt{u} \right)^2  \;  - \; 2 \; 
        \left( \arcsin \sqrt{\frac{u}{2}} \right)^2    \right)
      \label{r8}   \\
      P_5 (u)&=&  
      \frac{10}{\pi }\left( 
      \arcsin \sqrt{u}    \;  - \; 2 \;  \arcsin \left( \cos \frac{\pi }{5}
     \; \sqrt{u} \right)
      \;  + \; 2 \;  \arcsin \left( \cos \frac{2 \pi }{5}
     \; \sqrt{u} \right)
      \right)
      \label{r9}   
\eea

\noindent
These functions are displayed below in Figure 4.

\begin{figure}[!h]
\begin{center}
\includegraphics[scale=.5,angle=0]{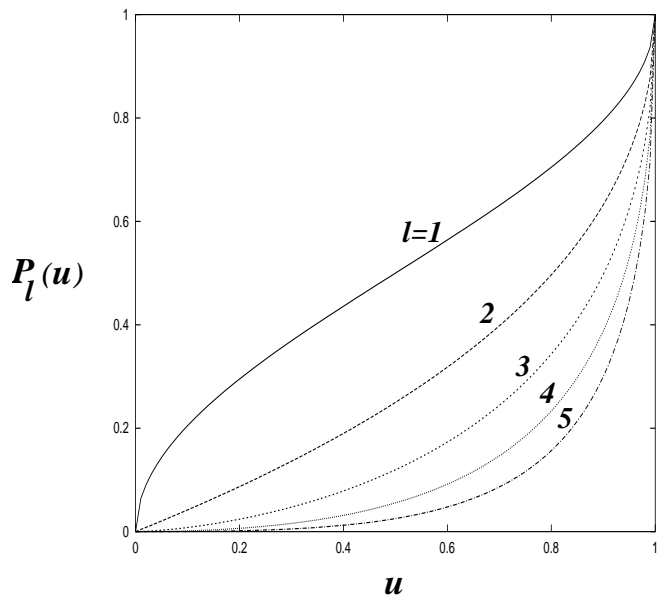}
\caption{ The distribution functions $P_l(u)$ for $l=1,\ldots ,5$. }
\end{center}
\end{figure}

\vskip.3cm
\noindent
As expected, the Levy's second arc-sine law is recovered in (\ref{r5}).
 Moreover, the result (\ref{r6}) is straightforward since, when $l=2$, the
  $2$ components of the Brownian motion factorize. Thus, for $l=2$,
   $(\arcsin )^2$ functions appear. What is surprising is that they will
    appear each time $l$ is even while being absent when $l$ is odd.
    
\vskip.3cm
\noindent
For the probability density, ${\cal P}_l $
($\equiv \frac{\d P_l}{\d u}$), with (\ref{r3}) and (\ref{r4}), we obtain:

\be\label{r10}
{\cal P}_l (u) \quad  \sim  \quad  \frac{l}{\pi} \; \frac{1}{\sqrt{1-u}}  
  \qquad   \mbox{ when }  \qquad  u \to 1^-
\ee

\vskip.3cm
\noindent
This is consistent with (\ref{n11}) that corresponds to $l=1$.

We now present a formula which relates the first passage and the last
passage time distribution. The starting point is (\ref{n1}) and (\ref{n2})
which may be rewritten as
\be\label{r11}
 P(g< u)=\int{ Pr(T> 1-u|{r_0} )
 \;  \frac{1}{u}  \: e^{- \frac{r_0^2}{2u}}\:{r_0} \:\d {r_0}}  
\ee
where $Pr(T> (1-u)|{r_0})$ is the probability distribution of the first
passage time $T$ through F, given that the process starts at $r_0$.
Then, by scaling one has
\be\label{r12}
 Pr(T> (1-u)|{r_0})= Pr(T> {(1-u)\over{r_0}^2}|1)
\ee
By a simple change of variables it follows that
\be\label{r13}
 P(g< {1\over1+t})=\int{ Pr(T> {t \over2x}|1 )
\; e^{-x} \:\d x}
\ee
Therefore
\be\label{r14}
 P(g< {1\over1+t})=
 E(e^{-{t\over 2T}})
\ee
which is a relation between the first passage characteristic function
for a process starting at $r_0=1$ and the probability distribution of the
last passage time. Interestingly enough this formula can also be derived in
a more intrinsic fashion   using only  time inversion and scaling
\cite {Y3}.
As an application, let us the derive the density of first passage time in a
wedge of angle $\phi= {\pi\over3 }$. 
  In the context of the capture problem mentioned in the
 introduction, this corresponds to 
 a set of three identical and independent particles \cite{NOC}.
 In this case,  the distribution $P(g)$ is given in 
eq (\ref{r7}). By an inverse Laplace transform  (\ref{r14}) gives the
density of first passage time:
\be\label{r15}
f(T)={ 6\over {\pi^{3\over2} T}}\; e^{-{1\over 2T}}\:(\:\int_{0}^{\sqrt
{1\over2T}} e^{y^2}\d y -2\int_{0}^{\sqrt
{1\over8T}} e^{y^2}\d y\;)
\ee
One can check that this formula is in agreement with  (16) of \cite{NOC}
which expresses the first collision time probability for a given set of
initial conditions. By averaging this formula over the angle
 and setting $r=1$ one recovers (\ref{r15}).

\vskip.3cm
\noindent
ACKNOWLEDGMENT
\noindent
We are particularly grateful to Prof. Marc Yor for introducing us to the
subject and for pointing out (\ref{r14}).

\end{document}